\documentclass[aps,8pt,prd,twocolumn,preprintnumbers,amsmath,amssymb,nofootinbib,superscriptaddress,a4paper,showpacs]{revtex4-1}

\usepackage[breaklinks]{hyperref}
\usepackage[english]{babel}
\usepackage{amsmath}
\usepackage{amsfonts,amsthm}
\usepackage{mathtools}
\usepackage[breaklinks]{hyperref}
\usepackage[english]{babel}
\usepackage{graphicx}
\usepackage{color}
\usepackage{bm,bbm}

\newtheorem{post}{Definition}

\newcommand{\al}{\ensuremath\alpha}
\newcommand{\be}{\ensuremath\beta}

\newcommand{\om}{\ensuremath\omega}
\newcommand{\Om}{\ensuremath\Omega}
\newcommand{\Ga}{\ensuremath\Gamma}

\newcommand{\beq}{\begin{equation}}
\newcommand{\eeq}{\end{equation}}
\newcommand{\bal}{\begin{eqnarray}}
\newcommand{\eal}{\end{eqnarray}}

\newcommand{\na}{\nabla}

\newcommand{\lr}[1]{\left(#1\right)}
\newcommand{\lrsq}[1]{\left[#1\right]}

\renewcommand{\bf}[1]{{\textbf{#1}}}
\newcommand{\bQ}{{\bf Q}}
\newcommand{\bom}{\bm{\omega}}
\newcommand{\bOm}{\bm{\Omega}}
\newcommand{\bg}{{\bf g}}

\newcommand{\dif}{\textbf{{d}}}

\begin{document}

\title{A generalized Weyl structure with arbitrary non-metricity}

\author{Adria Delhom}
\email{adria.delhom@uv.es}
\affiliation{Departamento de F\'{i}sica Te\'{o}rica and IFIC, Centro Mixto Universidad de
Valencia - CSIC. Universidad de Valencia, Burjassot-46100, Valencia, Spain}

\author{Iarley P. Lobo}
\email{iarley\_lobo@fisica.ufpb.br}
\affiliation{Departamento de F\'{\i}sica, Universidade Federal da Para\'iba\\ Caixa Postal 5008, 58051-970, Jo\~ao Pessoa, Para\'iba, Brazil}
\affiliation{Departamento de F\'isica, Universidade Federal de Lavras, Caixa Postal 3037, 37200-000 Lavras-MG, Brazil}

\author{Gonzalo J. Olmo}
\email{gonzalo.olmo@uv.es}
\affiliation{Departamento de F\'{i}sica Te\'{o}rica and IFIC, Centro Mixto Universidad de
Valencia - CSIC. Universidad de Valencia, Burjassot-46100, Valencia, Spain}

\author{Carlos Romero}
\email{cromero@fisica.ufpb.br}
\affiliation{Departamento de F\'{\i}sica, Universidade Federal da Para\'iba\\ Caixa Postal 5008, 58051-970, Jo\~ao Pessoa, Para\'iba, Brazil}

\begin{abstract}
A Weyl structure is usually defined by an equivalence class of pairs $(\bg, \bom)$ related by Weyl transformations, which preserve the relation $\na \bg=\bom\otimes\bg$, where $\bg$ and $\bom$ denote the metric tensor and a 1-form field. An equivalent way of defining such a structure is as an equivalence class of conformally related metrics with a unique affine connection $\Ga_{(\bm{\omega})}$, which is invariant under Weyl transformations. In a standard Weyl structure, this unique connection is assumed to be torsion-free and have vectorial non-metricity. This second view allows us to present two different generalizations of standard Weyl structures. The first one relies on conformal symmetry while allowing for a general non-metricity tensor, and the other comes from extending the symmetry to arbitrary (disformal) transformations of the metric. 
\end{abstract}


\maketitle

\section{Introduction}

General relativity and a major part of alternative theories of gravity are constructed on the assumption that spacetime is a differential manifold endowed with a metric and a Riemannian connection, i.e., a metric-compatible connection. This construction is fundamentally important in order to define scalars and a dynamics (from the notion of a derivative) that is manifestly covariant under general coordinate transformations, which encompasses the main symmetry presented in gravitational theories in the past 100 years.
\par
However, as already noticed by Weyl in \cite{weyl1}, Riemannian geometry is characterized by the invariance of lengths under parallel displacement. Although the direction of vectors may  get modified under parallel displacement due to a non-trivial curvature tensor, vector lengths are still preserved. Then, he demonstrated that indeed this was an assumption that, in principle, could be avoided when constructing a {\it true infinitesimal geometry}. Another key point for Weyl was that scalar products should be determined up to an arbitrary positive proportionality factor, i.e., the physics should be invariant up to a conformal transformation of the metric and only ratios would have physical meaning (angles are also naturally invariant). Also, if one uses only light signals to define coordinates of the spacetime manifold, such invariance would also be present since the line element is conformally invariant (as also noted by 't Hooft in  \cite{Hooft:2014daa}).
\par
Weyl presents two axioms for this geometry, the first one states that parallel displacement of vectors defines what he calls a {\it similarity map}, i.e., a linear map that induces a conformal transformation in the inner product of any two vectors. This axiom implies that for a given vector that is parallel displaced along any curve, the derivative of the vector's norm is proportional to the norm itself, and the proportionality function is given by some 1-form calculated along the curve, which is equivalent to require vectorial non-metricity of the form\footnote{The usual definition of the non-metricity tensor is $\bQ=\na\bg$.} $\bQ=\bom\otimes \bg$. The second axiom consists in the requirement of a torsion-free connection.
\par
It can be seen that a gauge transformation of this 1-form field can always compensate any conformal transformation of the metric. However, as it follows from the above considerations, the particular form of the non-metricity condition given by the first axiom is conceptually independent of the invariance of the physics under conformal transformations of the metric. This axiom for the non-metricity condition is just a matter of convenience, in which the rule that governs the norm of parallel displacement coincides with a conformal map of the norm.
\par
The reason for the conformal transformation of the metric has a strong motivation in Mach's principle. In fact, other authors refer to such physical invariance as a guidance towards deeper formulations of gravity, even in a Riemannian setting (see for instance \cite{Brans:1961sx,Dicke:1961gz,Mannheim:1988dj,Barbour:2010dp,Hooft:2014daa}).
\par
In this paper, we demonstrate that indeed it is possible to construct a geometry that is intrinsically invariant under a conformal transformation of the metric, with a gauge invariant connection for an arbitrary non-metricity, which naturally generalizes Weyl's construction. We also study the geometrical properties of a more general class of transformations, called disformal transformations, of which  conformal transformations are just a subclass. This kind of transformation has been extensively studied in several gravitational theories, for instance, in the original Bekenstein's approach \cite{beken1, beken2} and TeVeS formalism \cite{beken_mond} (which recovers MOND \cite{milgrom} in the nonrelativistic limit), scalar \cite{scalartheory} and scalar-tensor theories of gravity \cite{mota,ip,sak1,sak2} (like Horndeski's \cite{dario,miguel1,vernizzi} and Mimetic \cite{rua,matarrese,sunny1,sunny2,Sebastiani:2016ras} approaches), bimetric theories \cite{bimetric-clifton}, analogue models \cite{nov_bit_gordon,nov_bit_drag}, k-essence \cite{mukhanov}, cosmology \cite{nemanja,miguel2,bruck1,bruck2}, metric-affine gravity \cite{Afonso:2018bpv,BeltranJimenez:2017doy,Afonso:2017bxr} and quantum gravity phenomenology \cite{Olmo:2011sw}.
\par
The paper is organized as follows: in Section (\ref{sec:w-geom}) we review the basic concepts of Weyl geometry. In Section (\ref{sec:non-metric}) we give a generalization of a Weyl structure for general non-metricity tensors while maintaining conformal symmetry. We also present another possible generalization by extending conformal symmetry to the disformal case, which requires a completely general non-metricity tensor on its own. In Section (\ref{sec:appl}) we revisit the fundamental physical meanings of these transformations and demonstrate how the use of non-Riemannian geometries allows one to extend the usual diffeomorphism invariance to diffeomorphism+conformal /disformal symmetry providing explicit examples. We conclude in Section (\ref{sec:conc}).


\section{Revisiting Weyl geometry}\label{sec:w-geom}

In his seminal paper \cite{weyl1}, Weyl proposed two axioms for the geometrical setting that now goes under his name. They can be stated in a modern language by the following conditions:
\begin{post}
Let $M$ be a differentiable manifold \ endowed with
an affine connection $\Ga$, a metric tensor $\bg$, and a one-form field
$\bom$, called a Weyl field. It is said that $\Ga$ is compatible
($\bom$-compatible) with $\bg$ if for any vector fields $u,v,w\in T(M)$,\ we have%

\begin{equation}
\nabla_{v}\lrsq{g(u,w)}=g(\nabla_{v}u,w)+g(u,\nabla_{v}w)+\omega(v)g(u,w).
\label{W-compatible}%
\end{equation}
Besides that, for any $u,v\in T(M)$ the following condition holds:
\begin{equation}
\nabla_{v}u-\nabla_{u}v=[v,u]. \label{torsionless}
\end{equation}
We write $\Ga_{\bm{\omega}}$ for an $\omega$-compatible connection.
\end{post}

The first condition given by Eq.(\ref{W-compatible}) is the $\bom$-compatibility condition and determines the non-metricity tensor to be of the form $\bQ=\bom\otimes\bg$. The second condition, given by (\ref{torsionless}) states that the connection should be torsionless or symmetric.
\par
Weyl then realized that if one simultaneously performs a conformal transformation of the metric and a gauge transformation of the 1-form field of the form
\begin{equation}
\tilde{\bg}=e^{\phi}\bg\qquad\text{and}\qquad\tilde{\bom}=\bom+\dif \phi,\label{weyltransf}
\end{equation}
the compatibility condition remains invariant for any scalar function $\phi$. Such a pair of transformations of the metric and the 1-form field will be called Weyl transformation $W_\phi$. To verify the invariance of the compatibility condition under $W_\phi$, let us use the expression of the covariant derivative of the metric that can be deduced from the previous postulate: $\nabla\bg =\bom\otimes\bg$. This way, it becomes straightforward to check that $W_\phi\lrsq{\nabla\bg\otimes\bom}=\nabla\tilde{\bg}\otimes\tilde{\bom}$.  Thus, the equivalence class of pairs $(\bg,\bom)$ related by a Weyl transformation define a Weyl structure on $M$. 
\par
However, the invariance of the compatibility condition suggests that the affine connection remains invariant under a Weyl transformation. In order to verify this, let us point out that in any manifold with a metric $g$ and an affine connection $\Ga$, we can always make a decomposition of the connection symbols in a coordinate basis $\{\partial_{\mu}\}$  as
\begin{eqnarray}\label{conndecomp}
\Gamma^\al{}_{\mu\nu}=\left \{^{\al{}}_{\mu\nu}\right \}(\bg)-\frac{1}{2}\lr{Q_\mu{}^\al{}_\nu+Q_\nu{}^\al{}_\mu-Q^\al{} _{\mu\nu}}+K^\al{}_{\mu\nu}\;,
\end{eqnarray}
where $\left \{^{\al{}}_{\mu\nu}\right \}(\bg)$ are the Christoffel symbols of $\bg$ (which define the usual Levi-Civita connection of $\bg$), $K^\al{}_{\mu\nu}$ is the contortion tensor, which vanishes in the absence of torsion \cite{BeltranJimenez:2019tjy},  and $Q_{\alpha\mu\nu}=\na_\al g_{\mu\nu}$ are the components of the non-metricity tensor. Within this generic decomposition, since the compatibility condition \eqref{W-compatible} implies a non-metricity tensor given by $\bQ=\bom\otimes\bg$, the connection symbols of the unique affine connection defined by a Weyl structure in a coordinate basis read
 \beq\label{Weylconn}
\Gamma_{(\bom)}{}^\al{}_{\mu\nu}=\left \{^{\al{}}_{\mu\nu}\right \}(\bg)-\frac{1}{2}\lr{\om_\mu\delta^\al{}_\nu+\om_\nu\delta^\al{}_\mu-\om^\al g_{\mu\nu}}.
\eeq
This decomposition points to the fact that the 1-form, rather than being a fundamental object in a Weyl space, is a piece of its affine connection $\Ga_{\bm{\omega}}$. Since  the metric and the affine connection are two equally fundamental and independent geometrical entities,  transformations of the metric (for instance, a conformal transformation) should leave the connection unchanged. Therefore, in view of \eqref{Weylconn} , requiring the invariance of $\Gamma_{(\omega)}{}^\al{}_{\mu\nu}$ under a conformal transformation $\bg\mapsto e^\phi \bg$ implies the following condition
\begin{align*}
\delta\Gamma_{(\bm{\omega})}{}^\al{}_{\mu\nu}=&\lr{\partial_\mu\phi \,\delta^\al{}_\nu+\partial_\nu\phi\,\delta^\al{}_\mu-\partial^\alpha\phi\, g_{\mu\nu}}\\&-\lr{\delta\om_\mu\,\delta^\al{}_\nu+\delta\om_\nu\,\delta^\al{}_\mu-\delta\om^\al\, g_{\mu\nu}}=0,
\end{align*}
where $\delta\omega_\mu$ is the change induced in $\omega_{\mu}$ by the conformal transformation. This has a unique solution given by $\delta\bom=\dif\phi$, which leads to  the familiar gauge transformation of $\bom$ written in \eqref{weyltransf}.
\par
We have just seen that the independence between metric and connection is what ultimately allows us to recover the usual prescription for the change of $\bom$ under a conformal rescaling of the metric. In this light, a Weyl structure is an equivalence class of conformally related metrics in an affinely-connected manifold with the condition of having the non-metricity given by the one-form $\bom$. Its fundamental objects are, therefore, the conformal class of metrics $\lrsq{g}$ and the unique affine connection $\Gamma_{\bm{\omega}}$ defined by the one-form of the non-metricity condition. We can sum these conclusions up by giving a fundamental role to the conformal symmetry and the affine connection in the definition of a Weyl structure:
\begin{post}
A Weyl structure is a differentiable manifold $\mathcal{M}$ endowed with
a unique  torsion-free affine connection $\Ga_\omega$ and a conformally related equivalence class of metric tensors $\lrsq{g}$, such that the non-metricity tensor is given by $\bQ=\bom\otimes\bg$, where $\omega$ is any 1-form field and $\bg$ is a representative of the class.\label{def2}
\end{post}
Notice that this definition gives more emphasis to the conformal symmetry satisfied by the Weyl structure and the invariant elements within.\\
\par

Historically, the Weyl structure was related to the first attempt to define a unified field theory, in which gravity and electromagnetism (the forces known at the time) would be described by a single geometrical entity. In this case, the electromagnetic potential was intended to be the Weyl field due to its gauge symmetry, which is probably the reason why the 1-form field has been seen as a fundamental piece of Weyl structures, to the detriment of the invariant affine connection. In this unified picture of Weyl, electro-gravitational field equations would be derived from a $\sqrt{-g}R^2$ action, where $R$ is the curvature scalar of $\Ga_\omega$. This action is indeed invariant under the set of transformations \eqref{weyltransf} and represents one of the first proposals for a metric-affine $f(R)$ model. 
\par
Such unified formulation of gravity and electromagnetism was promptly criticized by Einstein in a postscript attached to the debut paper (\cite{weyl1}), which received a subsequent reply by Weyl (which can also be found in \cite{weyl1}), in which it is made clear that the behaviour of clocks and rods as stated by Einstein would need to be revisited in face of a more fundamental theory. 
\par
A renewed interest in alternative formulations of gravity in non-Riemannian geometries has emerged in recent years. These new theories do not pursue the idea of unifying gravity and other forces (which would require a satisfactory understanding of the quantum nature of the gravitational field, see for instance \cite{Harko:2018gxr,BeltranJimenez:2019tjy,deCesare:2016mml,Ghilencea:2018dqd, Romero:2012hs,Lobo:2015zaa,Avalos:2018uvq,Lobo:2018zrz,Scholz:2017pfo}). In the next section we will discuss how Weyl structures can be generalized for the case of a general non-metricity tensor.


\section{Conformal generalization of a Weyl structure}\label{sec:non-metric}

In the perspective outlined above, a Weyl structure has two main ingredients: conformal symmetry and vectorial non-metricity. In this sense, Weyl symmetry is nothing but conformal symmetry in an affinely connected space-time with vectorial non-metricity. Therefore, there are two ways of generalizing this structure: 1) dropping the vectorial non-metricity postulate or 2) changing the symmetry on which the structure is based. In this section we will generalize the standard Weyl structure by relaxing the vectorial non-metricity constraint while maintaining conformal symmetry. To that end, let us generalize the argument in (\ref{sec:w-geom}) without the vectorial non-metricity restriction. 
\par
Consider a manifold with a conformally related equivalence class of metrics $\lrsq{g}$ and a unique affine connection $\Ga$, which can always be decomposed as in \eqref{conndecomp}, where $\bg$ is a representative of the class. For simplicity, let us impose the torsion-free condition so that $K^\al{}_{\mu\nu}=0$ in the decomposition. Now from invariance of the affine connection under a conformal transformation $\bg\mapsto e^\phi \bg$, it follows that 
\begin{align*}
\delta\Gamma^\al{}_{\mu\nu}=&\lr{\partial_\mu\phi \,\delta^\al{}_\nu+\partial_\nu\phi\,\delta^\al{}_\mu-\partial^\alpha\phi\, g_{\mu\nu}}\\&-\lr{\delta Q_\mu{}^\al{}_\nu+\delta Q_\nu{}^\al{}_\mu-\delta Q^\al{} _{\mu\nu}}=0,
\end{align*}
which has as a unique solution $\delta Q_{\mu}{}^{\al}{}_{\nu}=\partial_\mu\phi \delta^\al{}_\nu$. This leads to a transformed non-metricity tensor that must be
\begin{equation}
\tilde{\bQ}=e^\phi\lr{\bQ+\dif\phi\otimes\bg}
\end{equation}
 Notice that the trace corresponding to the Weyl non-metricity transforms as $\tilde{Q}_{\mu\al}{}^{\al}=Q_{\mu\al}{}^{\al}+N\partial_\mu\phi$, where $N$ is the space-time dimension, recovering the standard gauge transformation for $Q_{\mu\al}{}^{\al}=N\omega_\mu$. The fact that a conformal transformation of the metric can be absorbed in one of the traces of the non-metricity tensor as a gauge transformation justifies Weyl's interpretation of the 1-form $\omega_\mu$ as the photon field. However, from the above analysis, we can see that the vectorial non-metricity condition is an ad-hoc condition that we have no physical reason to assume.  Dropping the postulate of vectorial non-metricity from Definition \ref{def2}, we are led to define the following generalized Weyl structure
\begin{post}
A conformally generalized Weyl structure is a differentiable manifold $\mathcal{M}$ endowed with
a unique torsion-free affine connection $\Ga$ and a conformally related equivalence class of metric tensors $\lrsq{g}$.\label{defgenWeyl}
\end{post}
Since this generalized Weyl structure is grounded on conformal symmetry, it will be useful to describe space-times with a non-Riemannian connection that exhibit conformal symmetry. The corresponding generalized Weyl transformation is actually nothing but a conformal transformation (as in the standard Weyl case). However, if we want to split the connection and view Weyl transformations in the traditional sense of conformal transformations of the metric plus a gauge transformations of some 1-form, we end up with the following generalized Weyl transformations
\begin{equation}
\tilde{\bg}=e^{\phi}\bg\qquad\text{and}\qquad\tilde{\bQ}=e^\phi \lr{\bQ+\dif \phi\otimes\bg}.
\label{genweyltransf}
\end{equation}
We see that analogously to the gauge transformation of $\bom$, the generalization to arbitrary non-metricity is standard, implying only a conformal transformation of the non-metricity tensor plus a gauge transformation of the corresponding trace. 


\subsection{On the physical equivalence of conformal frames}

The issue raised by Weyl about the physical equivalence of conformal metrics has been explicitly discussed in Brans-Dicke theory of gravity  \cite{Brans:1961sx}. This theory is one of the most fruitful attempts to extend general relativity by introducing a non-minimal coupling between a scalar field and the curvature. This construction allowes a conformal transformation of the metric to be compensated by a redefinition of this scalar field. This mapps a field theory with non-minimal coupling, but with a variable (effective) gravitational ``constant'' (the Jordan frame) to a theory that resembles general relativity with a scalar field but with the non-minimal coupling transferred to the matter sector.
\par
Specifically, this property was originally discussed in \cite{Dicke:1961gz}, where Dicke described how his scalar-tensor theory could incorporate an invariance under transformation of units, besides the usual coordinate invariance. In that paper, Dicke specifically stated that he would only analyze conformal transformations as a particular case of such ``transformation of units'', but that in principle a more general case could be considered.
\par
Brans-Dicke theory was originally intended to be a gravitational theory closer to Mach's principle than general relativity. In fact the Machian nature of this theory manifests itself by the existence of this functional dependence of the gravitational ``constant'' ($G$), which means that the coupling parameter that governs the gravitational interaction would depend on the mass distribution in the universe. As said above, it turns out that it was also realized the important role of the conformal symmetry, since this formalism allowed one to expand the symmetry principle of gravitational theories beyond the usual coordinate transformation in such a way that it is always possible to move from a frame in which $G$ is variable to one in which $G$ is a true constant.
\par
Complementarily, Mach's principle has been reformulated in order to include the conformal symmetry in recent years \cite{Barbour:2010dp}. Therefore, since in Brans-Dicke theory it is always possible to have $G$ constant in a frame, the Machian character of this theory seems to be more transparently manifest by the conformal symmetry rather than by the variable-$G$.
\par
On the other hand, it has been discussed in the literature that Brans-Dicke theory finds a natural language in Weyl geometry \cite{Quiros:2011wb,Lobo:2016izs,Quiros:2018ryt,Quiros:2019ktw}, since Weyl's non-metricity has always been regarded as the one in which conformal symmetry is naturally incorporated. As we have shown that for arbitrary non-metricity tensors it is also possible to formulate a conformal invariance principle, the existing link between the Brans-Dicke theory and Weyl geometry means that it is possible to generalize Brans-Dicke theory without loosing its Machian motivation by considering more general non-metricity tensors, which is a nontrivial result that follows from our analysis. 
\par
As an example, consider the action 
\begin{equation}\label{exp1}
S=\int d^4x\sqrt{-g}\left[\alpha \left(g^{\mu\nu}R_{\mu\nu}\right)^2+\beta R^{\mu\nu}R_{\mu\nu}\right],
\end{equation}
in which metric and connection are regarded as independent fields, $R_{\mu\nu}=R_{\mu\nu}(\Gamma)$, and $\alpha$ and $\beta$ are constants. It is easy to verify that this theory is conformally invariant. {The addition of matter could be done in a non-invariant way by means of a matter Lagrangian defined by the usual minimal coupling, i.e., by promoting the Minkowski metric to a general curved metric and partial derivatives to covariant ones. This is what is done in Brans-Dicke gravity, which formally breaks the conformal invariance, but raises a discussion about the physical equivalence of the different conformal frames (like Jordan and Einstein ones) \cite{Quiros:2011wb}. The point of view that treats these frames as equivalent argues that although the forms of the field equations are different, a conformal transformation in the metric would not modify physical observables, since they are defined by ratios and angles.}
Another possibility consists in following a procedure similar to what is done in \cite{Almeida:2013dba} in the context of a Geometrical Brans-Dicke theory, in which the matter coupling is done with a symmetric $(0,2)$-tensor (an effective metric) that is invariant under Weyl transformations. In this case, one defines the matter Lagrangian with the effective metric $\gamma_{\mu\nu}=e^{-\phi}g_{\mu\nu}$ (where $\phi$ plays the role of the Brans-Dicke or Weyl field) and the energy-momentum tensor from a variation with respect to $\gamma_{\mu\nu}$. This way, the matter sector becomes naturally invariant under Weyl transformations. For the case of a general non-metricity, one could use a function with a $-1$ conformal weight
\begin{equation}
h(e^{f}g,\Gamma)=e^{-f}\, h(g,\Gamma),
\end{equation}
in order to define an effective metric $\gamma_{\mu\nu}=h(g,\Gamma)g_{\mu\nu}$.
\par
For example, a function like $h(g,\Gamma)=R(g,\Gamma)/\Lambda^2$, where $R(g,\Gamma)$ is the curvature scalar and $\Lambda$ is a constant that defines an energy-scale, might serve our purposes. A natural consequence of this coupling is that particles would follow riemannian geodesics of the effective metric $\gamma_{\mu\nu}$ in any conformal gauge, thus assuring the invariance principle discussed throughout this paper. A similar non-minimal coupling was also proposed in \cite{Gomes:2018sbf} in the context of Weyl gravity.


\section{Disformal generalization of a Weyl structure}\label{sec:appl}

As explained above, besides dropping the vectorial non-metricity condition, another possible generalization of a Weyl structure goes by changing the underlying symmetry principle. The most conservative generalization in this direction would be to extend the symmetry group while keeping the vectorial non-metricity postulate. Therefore, the first thing we should do in this regard is to understand whether this can be done or, rather, if we are forced to drop the vectorial non-metricity postulate when enhancing the symmetry.
\par
To that end, let us find out what are the most general transformations of the metric that keep the vectorial character of non-metricity unchanged. From the definition of the non-metricity tensor and invariance of the affine connection,  if we start with vectorial (or Weyl-like) non-metricity $\bQ= \bom\otimes\bg $, after a general transformation of the metric like 
\begin{equation}
\tilde{g}_{\mu\nu}=g_{\al\be}\Om^\al{}_\mu\Sigma^\be{}_\nu,\label{disformaltrans}
\end{equation}
where $\Omega^\al{}_\be$ and $\Sigma^\al{}_\be$ are invertible, we are led to a transformed non-metricity of the form
\beq
\tilde{Q}_{\rho\mu\nu}=\om_\rho \tilde g_{\mu\nu}+g_{\mu\be}\na_\rho\lr{\Om^\al{}_\mu\Sigma^\be{}_\nu}.
\eeq
By demanding that the transformed non-metricity be also of a Weyl-like form,  we see that there must exist a 1-form $\tilde\bom$ satisfying
$$\tilde\om_\rho \tilde{g}_{\mu\nu}=\om_\rho \tilde g_{\mu\nu}+g_{\al\be}\na_\rho\lr{\Om^\al{}_\mu\Sigma^\be{}_\nu},$$
 which can be satisfied only if  $g_{\al\be}\na_\rho\lr{\Om^\al{}_\mu\Sigma^\be{}_\nu}=\xi_\rho\tilde g_{\mu\nu}$ and therefore, that there exists a  1-form $\boldsymbol{\xi}$ such that $$\na_\rho\lr{\Om^\al{}_\mu\Sigma^\be{}_\nu}=\xi_\rho \Om^\al{}_\mu\Sigma^\be{}_\nu.$$ 
Hence we would have
\begin{align*}\label{eq-xi}
&\tilde{\omega}_{\mu}=\omega_{\mu}+\xi_{\mu},\\
&\xi_{\mu}=\frac{1}{N}\lr{(\Omega^{-1})^{\al}{}_{\beta}\nabla_{\mu}\Omega^{\beta}{}_{\al}+(\Sigma^{-1})^{\al}{}_{\beta}\nabla_{\mu}\Sigma^{\beta}{}_{\al}},
\end{align*}
where $N$ is the space-time dimension. As a matter of fact, this expression can be simplified by explicitly writing the covariant derivative of $\xi$ in terms of its partial derivative and the connection symbols. The contributions from contravariant and covariant indices are cancelled out and we end up with 
\beq
\xi_{\mu}=\frac{1}{N}\lr{(\Omega^{-1})^\al{}_{\beta}\, \partial_{\mu}\Omega^{\beta}{}_{\al}+(\Sigma^{-1})^\al{}_{\beta}\, \partial_{\mu}\Sigma^{\beta}{}_{\al}}\label{eqxi}.
\eeq
By computing $d\xi$ in terms of $\bOm$ and its partial derivatives from \eqref{eqxi} we find
\begin{equation}\label{rot-xi1}
\begin{split}
\partial_{\mu}\xi_{\nu}-\partial_{\nu}\xi_{\mu}=\frac{1}{d}\left[\partial_{\mu}(\Omega^{-1})^{\alpha}{}_{\beta}\, \partial_{\nu}\Omega^{\beta}{}_{\alpha}-\partial_{\nu}(\Omega^{-1})^{\alpha}{}_{\beta}\, \partial_{\mu}\Omega^{\beta}{}_{\alpha}\right.\\
+\left. \partial_{\mu}(\Sigma^{-1})^{\alpha}{}_{\beta}\, \partial_{\nu}\Sigma^{\beta}{}_{\alpha}-\partial_{\nu}(\Sigma^{-1})^{\alpha}{}_{\beta}\, \partial_{\mu}\Sigma^{\beta}{}_{\alpha}\right].
\end{split}
\end{equation}
By using $\partial_{\mu}\left((\Omega^{-1})^{\alpha}{}_{\beta}\Omega^{\beta}{}_{\rho}\right)=0$, we can derive the following identity: $\partial_{\mu}(\Omega^{-1})^{\alpha}{}_{\sigma}=-(\Omega^{-1})^{\alpha}{}_{\beta}\left(\partial_{\mu}\Omega^{\beta}{}_{\rho}\right)(\Omega^{-1})^{\rho}{}_{\sigma}$, and the same for the $\Sigma$ term. Substituting this identity into (\ref{rot-xi1}), we arrive to
\begin{equation}
\partial_{\mu}\xi_{\nu}-\partial_{\nu}\xi_{\mu}=0,
\end{equation}
i.e, the 1-form $\xi$ is integrable, and by the Poincar\'e lemma,this means that there exists a scalar function $\phi$ such that $\xi=\dif\phi$, and therefore $\na_\al\lr{\Omega^\mu{}_\nu\Sigma^\rho{}_\sigma}=\lr{\na_\al\phi}\Omega^\mu{}_\nu\Sigma^\rho{}_\sigma$. Without adding  extra structure, the only (1,1) tensor which is proportional to itself under covariant differentiation is the identity tensor $\delta$, and therefore the solution to this equation implies that $\Omega^\al{}_\be$ and $\Sigma^\al{}_\be$ must be given by
\begin{equation}
\Omega^{\alpha}{}_{\beta}=e^{A}\delta^{\alpha}_{\beta}\quad\text{and}\quad\Sigma^{\alpha}{}_{\beta}=e^{B}\delta^{\alpha}_{\beta},
\end{equation}
where $A$ and $B$ are scalar functions satisfying $A+B=\phi$. This shows that the most general transformation of the metric that leaves the connection invariant but preserves the vectorial (Weyl-like) character of non-metricity is a conformal transformation. Therefore, we can only consider a generalization of a Weyl structure to disformal symmetry if we also abandon the vectorial non-metricity condition. The quick way to understand this is as follows: given the Weyl connection (\ref{Weylconn}), we cannot ask for the Weyl structure to be invariant under more general transformations of the metric than conformal, since this would necessarily change the Levi-Civita part of (\ref{Weylconn}) in a way that can not be absorbed entirely in the $\bom$ 1-form, therefore shifting away from the connection \eqref{Weylconn}, which is by construction contradictory (keep in mind that we are only doing transformations of the metric). This signals that the fundamental property of the  standard Weyl structures, rather than the kind of non-metricity, is the symmetry group, which makes the vectorial non-metricity postulate seem rather artificial and further motivates us to consider the conformal generalization to the standard Weyl structure in section \ref{sec:non-metric}.\\

\subsection{Two examples of disformal Weyl structures.}
 Once understood that disformal generalizations of a Weyl structure necessarily lead to abandoning the vectorial non-metricity postulate, let us consider two possible kinds of such generalizations: one relying on $GL\lr{N,\mathbb{R}}$ as the symmetry group, and another kind of deformations which are present in the so-called Ricci-Based Gravity theories (RBGs), a broad class of metric-affine theories with projective symmetry recently considered in the literature \cite{BeltranJimenez:2017doy,Afonso:2017bxr,Afonso:2018bpv,Afonso:2018mxn,Afonso:2018hyj,BeltranJimenez:2019acz,Afonso:2019fzv}.
\par
The first kind consists on enhancing the symmetry group from the conformal group to $GL\lr{N,\mathbb{R}}$. The action of the general linear group on the metric is given by \cite{Capozziello:2011et}
\begin{equation}
\tilde{g}_{\mu\nu}=g_{\al\be}\Om^\al{}_\mu\Om^\be{}_\nu,\label{disformaltrans}
\end{equation}
where $\bOm$ is a rank (1,1) tensor with non-zero determinant (i.e. an element of the general linear group of the corresponding tangent space). Notice that, like conformal transformations, this is a transformation of the metric alone that does not affect the connection. By decomposing the affine connection and requiring invariance, we arrive at the corresponding transformation law that must be satisfied by the non-metricity tensor after a transformation of the metric like \eqref{disformaltrans}, which is
\begin{equation}
\tilde Q_{\rho\mu\nu}=Q_{\rho\al\be}\Om^\al{}_\mu \Om^\be{}_\nu +g_{\al\be}\na_\rho\left(\Om^\al{}_\mu\Om^\be{}_\nu\right).
\end{equation}

 
The second kind of disformal transformations are those inspired by RBGs  \cite{BeltranJimenez:2019acz,BeltranJimenez:2017doy,Afonso:2017bxr}, in which the space-time metric is related to an auxiliary metric associated to the independent affine connection in the following way:
\begin{equation}
 \tilde{g}_{\mu\nu}=g_{\mu\al}\Om^\al{}_\nu,\label{RBGdisf}
\end{equation}
where $\Om^\al{}_\nu$ is an invertible matrix. In analogy with RBGs,  $\Omega_{\al\be}$ must be symmetric when the upper index is lowered with any of the metrics. 
 Requiring the invariance of the affine connection under the above transformation \eqref{RBGdisf} leads to 
\begin{equation}\label{transf-q}
\tilde{Q}_{\al\mu\nu}=Q_{\al\mu\be}\Om^\be{}_\nu +g_{\mu\be}\na_\al\Om^\be{}_\nu.
\end{equation}


\subsection{Examples of disformal transformations and the frame equivalence issue}

As mentioned above, disformal transformations arise naturally in metric-affine gravity theories of the RBG type, such as $f(R,R_{\mu\nu}R^{\mu\nu})$ or Born-Infeld inspired gravity theories \cite{BeltranJimenez:2017doy,Olmo:2011sw}. {Recently, the non-metricity tensor induced by such disformal transformation has been observationally constrained in \cite{Latorre:2017uve}.}
\par
{The issue of the disformal invariance of cosmological observables has been treated in \cite{Domenech:2015hka} for a disformal transformation of the type}
\begin{equation}\label{exp2}
\Omega^{\alpha}{}_{\nu}=A\, \delta^{\alpha}_{\nu}+B\, \partial^{\alpha}\phi\partial_{\nu}\phi,
\end{equation}
where $A=A(x)$, $B=B(x)$ and $\phi=\phi(x)$ are scalar functions. In this case, not only the causal structure is preserved in cosmology but also the observables are invariant.
\par
A similar transformation is considered in \cite{Galtsov:2018xuc} in the Palatini formalism, in that case, there are some situations in which the disformal vector $\partial_{\mu}\phi$ is null-like \cite{Lobo:2017bfh}, which implies that gravitational waves propagate at the speed of light, thus preserving the causal structure.
\par
The issue of the disformal invariance of matter fields has been analyzed for the scalar \cite{Falciano:2011rf}, electromagnetic \cite{Goulart:2013laa} and spinor \cite{Bittencourt:2015ypa} cases. The scalar case considers a transformation of the kind (\ref{exp2}), the electromagnetic one considers
\begin{equation}\label{exp3}
\Omega^{\alpha}{}_{\nu}=A\, \delta^{\alpha}_{\nu}+B\, F^{\alpha\beta}F_{\beta\nu},
\end{equation}
where $F_{\alpha\beta}=\partial_{\alpha}A_{\beta}-\partial_{\beta}A_{\alpha}$ is the Maxwell tensor of the electromagnetic field $A_{\mu}$. For spinors, the transformation is of the form
\begin{equation}\label{exp4}
\Omega^{\alpha}{}_{\nu}=A\, \delta^{\alpha}_{\nu}+B\, J^{\alpha}J_{\nu}+C\, I^{\alpha}I_{\nu}+D\, J^{\left(\alpha\right.}I_{\left.\nu\right)},
\end{equation}
where $J^{\mu}\doteq\bar{\Psi}\gamma^{\mu}\Psi$ is the Dirac current, and $I^{\mu}\doteq\bar{\Psi}\gamma^{\mu}\gamma_5 \Psi$ is the axial current of the spinor field $\Psi$. As before $A,\, B,\, C$ and $D$ are scalar functions and parenthesis means symmetrization of indices.
\par
The issue of the disformal invariance was discussed by finding conditions on the scalar functions $A$-$D$ such that the Klein-Gordon, Maxwell and Dirac equations were kept invariant under the transformations (\ref{exp2}), (\ref{exp3}) and (\ref{exp4}), assuming a Riemannian manifold. 
\par
However, as a consequence of our discussion, the use of a non-Riemannian geometry with a connection that presents a non-metricity tensor allows us to face this issue from a different perspective by the introduction of new degrees of freedom that lead to disformal invariance. To exemplify, let us consider first the case of a massless scalar field that obeys the Klein-Gordon equation with a non-metric connection
\begin{equation}
\tilde{g}^{\mu\nu}\tilde{\nabla}_{\mu}\tilde{\nabla}_{\nu}\phi=0.
\end{equation}
\par
From the transformation rules (\ref{RBGdisf}) and (\ref{exp2}), we can deduce the disformal inverse metric as $\tilde{g}^{\mu\nu}=\Delta^{\mu}{}_{\alpha}g^{\alpha\nu}$, where $\Delta^{\mu}{}_{\alpha}\doteq C\delta^{\mu}_{\alpha}+D\partial^{\mu}\phi\partial_{\alpha}\phi$ corresponds to the same disformal rule (\ref{exp2}) of the covariant components of the metric, according to the redefinition $C=A^{-1}$ and $D=-B/A(A+w B)$, where $w=\partial^{\mu}\phi\partial_{\mu}\phi$.
\par
We demonstrated that if we make the simultaneous transformation (\ref{RBGdisf}) and (\ref{transf-q}), the connection remains invariant, consequently we naturally have that $\tilde{\nabla}_{\mu}\tilde{\nabla}_{\nu}\phi=\nabla_{\mu}\nabla_{\nu}\phi$. Besides that, a straightforward calculation implies that
\begin{eqnarray}
\tilde{g}^{\mu\nu}\tilde{\nabla}_{\mu}\tilde{\nabla}_{\nu}\phi=A^{-1}g^{\mu\nu}\nabla_{\mu}\nabla_{\nu}\phi\nonumber \\
-\frac{B}{A(A+wB)}\left(\frac{1}{2}\partial^{\mu}\phi\partial_{\mu}w-Q_{\mu\nu\alpha}\partial^{\mu}\phi\partial^{\nu}\phi\partial^{\alpha}\phi\right),
\end{eqnarray}
Therefore for a non-metricity tensor like
\begin{equation}
Q_{\alpha\mu\nu}=\frac{\partial^{\beta}\phi\partial_{\beta}w}{2w^3}\partial_{\alpha}\phi\partial_{\mu}\phi\partial_{\nu}\phi,
\end{equation}
we can map $\tilde{g}^{\mu\nu}\tilde{\nabla}_{\mu}\tilde{\nabla}_{\nu}\phi=0 \mapsto g^{\mu\nu}\nabla_{\mu}\nabla_{\nu}\phi=0$. For the massive Klein-Gordon equation, besides these considerations, a usual scaling of the field's mass like $m^2\mapsto A^{-1}\, m^2$ preserves the disformal invariance. The invariance of the other fields is more subtle and shall require a specific analysis elsewhere.
\par
Another context in which the disformal symmetry could be of use is in the formulation of purely affine theories of gravitation (see e.g. \cite{Einsteinaffine1,Einsteinaffine2,Eddingtonaffine1,Schrodingeraffine1,Schrodingeraffine2,Kijowskiaffine1,Kijowskiaffine2,Kijowskiaffine3,Poplawski:2007ik,Castillo-Felisola:2015cqa,Kijowski:2018onf}). Since it is easier to construct kinetic terms for matter fields with the help of a metric, purely affine theories could make use of disformal symmetry in order to introduce a fiducial metric. Given that disformal symmetry implies that  physical observables are invariant under arbitrary changes of the metric, this metric would not play any physical role, but might indeed help with the construction of the theories.
\par
For instance, consider an action of the kind
\begin{equation}
S=\int d^4x \mathfrak{L}(\Gamma^{\alpha}_{\mu\nu},R^{\mu}{}_{\alpha\beta\gamma},R_{\mu\nu},...),
\end{equation}
where $\mathfrak{L}$ is a Langrangian density which is a function of the connection, its curvature tensor, its Ricci tensor and possibly other fields (for instance a Curtright field in \cite{Castillo-Felisola:2015cqa}). Although this Lagrangian does not depend on the spacetime metric, it is possible to define a rank-two tensor density
\begin{equation}\label{f-metric}
\frac{\delta}{\delta R_{(\mu\nu)}}S[\Gamma]\doteq \sqrt{-g}g^{\mu\nu},
\end{equation}
which coincides with the usual metric density if we have the Einstein-Hilbert action. Notice that due to the invariance of the connection, the auxiliary tensor density (\ref{f-metric}) is preserved under the transformations that we present in this paper.


\section{Summary and conclusion}\label{sec:conc}

Weyl geometry has been faced as a natural arena in which conformal invariance (along with gauge invariance) can be realized. We have seen that theories constructed with a non-metric connection and having zero conformal weight are candidates of field theories that manifest such invariance. In this paper, we have revisited Weyl's original approach showing how even for general non-metricity tensors, we still have an underlying conformal symmetry principle. 
\par
Since the metric and the connection are two independent geometrical objects, a transformation of the metric should not affect the connection unless a rigid rule connecting them is assumed (like for instance the Riemannian compatibility condition). Hence, a transformation of the metric alone always introduces a corresponding transformation of the non-metricity tensor such that the connection remains invariant. Particularly, the Weyl group is just a special case in which the transformation of the metric is a conformal one and the non-metricity is vectorial.
\par
Furthermore, we have also analyzed the more general case of disformal transformations, and computed the change undergone by the non-metricity tensor so that invariance of the connection is preserved. As a corollary of this result, we showed that without adding extra structure, the most general kind of transformations of the metric that preserves the vectorial (or Weyl-like) character of non-metricity are conformal transformations. 
\par
Finally we discussed the physical meaning of conformal/disformal invariance as a symmetry property that extends the usual diffeomorphism invariance of gravitational theories, and its effect on the disformal transformations of a scalar field. The analysis of particular theories and physical imprints of such invariance principle will be the subject of future work.

\section*{Acknowledgments}
A. D. is supported by a PhD contract of the program FPU 2015 (Spanish Ministry of Economy and Competitiveness) with reference FPU15/05406. G. J. O. is funded by the Ramon y Cajal contract RYC-2013-13019 (Spain).  This work is supported by the Spanish projects FIS2014-57387-C3-1-P  and FIS2017-84440-C2-1-P (MINECO/FEDER, EU), the project SEJI/2017/042 (Generalitat Valenciana), the project i-LINK1215 (A.E. CSIC), the Consolider Program CPANPHY-1205388, and the Severo Ochoa grant SEV-2014-0398 (Spain). I. P. L. and C. R. thank Conselho Nacional de Desenvolvimento Cient\'ifico e Tecnol\'ogico (CNPq) and Coordena\c c\~ao de Aperfei\c coamento de Pessoal de N\'ivel Superior (CAPES) - Finance Code 001, from Brazil, for financial support.

\end{document}